# Giant Electrostatic Modification of Magnetism *via* Electrolyte-Gate-Induced Cluster Percolation in $La_{1-x}Sr_xCoO_{3-\delta}$

Jeff Walter,[1] T. Charlton,[2] H. Ambaye,[2] M.R. Fitzsimmons,[2,3] Peter P. Orth,[4]
R.M. Fernandes[5] and Chris Leighton[1]*

[1]Department of Chemical Engineering and Materials Science,
University of Minnesota, Minneapolis, MN 55455, USA
[2]Neutron Scattering Division, Oak Ridge National Lab, Oak Ridge, TN 37830, USA
[3]Department of Physics and Astronomy, University of Tennessee, Knoxville, TN 37996, USA
[4]Department of Physics and Astronomy, Iowa State University, Ames, IA 52242, USA
[5]School of Physics and Astronomy, University of Minnesota, Minneapolis, MN 55455, USA

**Abstract:** Electrical control of magnetism is a long-standing goal in physics and technology, recently developed electrolyte gating techniques providing a promising route to realization. Validating a recent theoretical prediction, here we demonstrate large enhancement of electrostatic modulation of ferromagnetic order in ion-gel-gated ultrathin $La_{0.5}Sr_{0.5}CoO_{3-\delta}$ by thickness-tuning to the brink of a magnetic percolation transition. Application of only 3-4 V then drives a transition from a short-range-ordered insulator to a robust long-range ferromagnetic metal, realizing giant electrostatic Curie temperature modulation over a 150 K window. *In operando* polarized neutron reflectometry confirms gate-controlled ferromagnetism, also demonstrating unusually deep penetration of induced magnetization, in further agreement with theory.

*Corresponding author: leighton@umn.edu

**PHYSH:** Physical Systems: Field-effect transistors, Oxides

Transport Properties: Percolation

Techniques: Neutron reflectometry

The electric field effect, wherein electric fields applied across a dielectric induce a controlled density of two-dimensional (2D) charge carriers in a conductor, has played a major role in science and technology. This is the principle of operation of the metal-oxide-semiconductor field-effect transistor (MOSFET), also being widely used in research for electrostatic gating [1-3]. Such devices are limited to modest charge densities ($10^{13}$ cm$^{-2}$ with SiO$_2$ dielectrics [1-3]), however, making electric double layer transistors (EDLTs) an exciting development [4-30]. In EDLTs the dielectric is replaced with an electrolyte such as an ionic liquid or gel, EDL formation generating massive capacitance, up to 100 µF cm$^{-2}$ [4-6]. This induces carrier densities up to $10^{15}$ cm$^{-2}$, 100 times higher than SiO$_2$ MOSFETs, and substantial fractions of an electron or hole per unit cell (u.c.) in most materials [1-3]. A new regime is entered at these densities where electronic phase transitions can be controlled. Superconductivity has been thus induced [7-11] (even discovered [8]) in insulating oxides (down to single u.c. thickness [9]), superconducting domes have been mapped [10-11], and insulator-metal transitions (IMTs) have been controlled [12-15].

Despite obvious potential, less progress has been made using EDLTs to control *magnetism* [16,17]. This is a long-standing challenge in physics and technology, as voltage-control of magnetic order and properties would provide many opportunities in data storage and processing [31-34]. As studies of electrolyte-gate-control of magnetism have expanded, a logical first step is control of the Curie temperature ($T_C$) in ferromagnetic (FM) conductors, *i.e.*, systems where magnetism and carrier density are coupled. In the last decade, the electrically-induced $T_C$ shift in EDLTs has risen from 30 K in La$_{0.8}$Ca$_{0.2}$MnO$_3$ [17] and SrRuO$_3$ [18], to 66 K in La$_{0.6}$Sr$_{0.4}$MnO$_3$ [19], 90 K in La$_{0.6}$Sr$_{0.4}$MnO$_3$/SrCoO$_{3-\delta}$ [20], 110 K in ultrathin Co [21], 130 K in Pr$_{0.55}$(Ca$_{0.7}$Sr$_{0.3}$)$_{0.45}$MnO$_3$ [22], and, recently, 200-225 K in La$_{0.5}$Sr$_{0.5}$CoO$_{3-\delta}$ [23] and (H)SrCoO$_{3-\delta}$ [24]. While this is impressive, it is vital to distinguish between *electrostatic* and *electrochemical* control [14,15,18,20,23-30]. Both approaches are of interest, but the additional ionic motion in electrochemical control could lead to slower, less reversible operation. In oxide EDLTs for example, field-induced O vacancy ($V_O$) creation and diffusion is now established [14,15,18,20,23-29], along with H injection and extraction [24,30], and it is exactly such "magneto-ionic" mechanisms that are implicated in the large $T_C$ shifts above [17-24]. Our own work on $La_{0.5}Sr_{0.5}CoO_{3-\delta}$ EDLTs provided an illustrative example by distinguishing electrochemistry at positive gate voltage ($V_g$) from predominantly electrostatic response at negative $V_g$ [23,25]. In essence, positive $V_g$ results in field-assisted $V_O$ creation [35] and diffusion, favored by low formation enthalpy. At negative $V_g$, however, annihilation of $V_O$ is thermodynamically disfavored, and electrostatic hole accumulation dominates. Electrostatic *vs.* electrochemical response is therefore understood based on $V_g$ polarity and the formation enthalpy and diffusivity of $V_O$ [23,25]. Critically, *electrochemical* control at $V_g > 0$ resulted in $La_{0.5}Sr_{0.5}CoO_3$ $T_C$ shifts of ~200 K in + 3 V [23], while *electrostatic* operation at $V_g < 0$ resulted in a $T_C$ shift of only 12 K in -4 V [25].

A natural question is thus how *electrostatic* control of magnetic order in such materials can be optimized. Importantly, many of these materials ($La_{1-x}Sr_xCoO_{3-\delta}$ (LSCO) [36-38], $La_{1-x}Sr_xMnO_3$ [39], *etc.*), evolve from inhomogeneous states to uniform FMs with doping. In LSCO for example, $Sr^{2+}$ creates formally $Co^{4+}$ ions that nucleate hole-rich nanoscopic FM clusters in an insulating non-FM matrix, eventually percolating into a long-range FM metal at $x_c = 0.18$ [36-38]. One attractive concept (Fig. 1(a)) is then to chemically dope to the brink of a percolation IMT (note the finite clusters (green) in Fig. 1(a)) and then electrostatically gate across the transition, potentially generating anomalously large increases in $T_C$, magnetization, and conductivity. This combined bulk chemical and surface electrostatic doping was considered in our recent percolation theory [40], resulting in Fig. 1(b). Solid lines here show the 2D surface charge densities (per Co) required to achieve percolation ($\Delta s_c$) *vs.* the starting effective chemical doping ($x_{eff}$), for multiple thicknesses (*t*). (We use $x_{eff}$ here due to finite $V_O$ concentration in $La_{1-x}Sr_xCoO_{3-\delta}$, which compensates Sr doping; in the simplest picture, $x_{eff} = x - 2\delta$ [41]). Independent of thickness, $\Delta s_c$ at $x_{eff} = 0$ is 0.5, the expected 2D value [42]. As $x_{eff}$ increases, $\Delta s_c$ first decreases linearly before dropping rapidly as the (*t*-dependent) bulk percolation threshold is approached [42]. Considering that an experimentally achievable $\Delta s$ in a perovskite EDLT is ~0.1 (shaded region, Fig. 1(b)), the steepness of $\Delta s_c$ near 3D bulk percolation (black line) means that tuning $x_{eff}$ to the brink of percolation in thick films would require unreasonable compositional control. As also shown in Fig. 1(b), however, $\Delta s_c(x_{eff})$ at low thickness shows progressively shallower approach to percolation. At 2 u.c., for example, an achievable $\Delta s \approx 0.1$ enables percolation at $0.23 < x_{eff} < 0.27$, *a 400 times wider window than 3D*. Thickness tuning to the brink of percolation followed by electrolyte gating is thus predicted as a promising means to optimize electrostatic control of magnetism.

Here, we first validate these predictions through transport studies of LSCO EDLTs *vs.* $x_{eff}$ and thickness ($t_{exp}$). Thickness tuning is indeed established as an ideal means to tune to the brink of a percolation IMT, 6 u.c. proving optimal. Ion gel gating of 6 u.c. films of $x = 0.5$ LSCO is then shown to enable *electrostatic* tuning from a short-range-ordered insulator to a long-range FM metal, spanning a 150 K $T_C$ range with only -4 V. The induced FM is robust, with 1 T coercivity, high remanance, and perpendicular anisotropy. *In operando* polarized neutron reflectometry (PNR) confirms FM, also establishing deeper penetration of induced magnetization than naively expected, in further agreement with theory.

Epitaxial LSCO EDLTs utilizing solid-state ion gel electrolytes based on 1-ethyl-3-methylimidazolium bis(trifluoro-methylsulfonyl) imide ionic liquid were prepared and characterized [23,25], as detailed in Supplemental Material Sec. A (Fig. S1) [43]. *Throughout this* *paper, only negative $V_g$s are applied, utilizing electrostatic (not electrochemical) gating*. Transport and PNR details are provided in Supplemental Material Sec. B [43]. Fig. 1(c-e) first shows the effect of varying $x_{eff}$ while keeping $t_{exp}$ approximately constant at 8-12 u.c., *i.e.,* the thick-film-limit in this system, where dead layer effects are weak [44]. Starting at $x = 0.50$ (Fig. 1(e)), as in prior work [25], the temperature ($T$) dependence of resistivity ($\rho$) [45] displays clearly metallic behavior, well beyond the percolation IMT. Applying $V_g = -4$ V decreases the low $T$ resistivity by ~18% *via* hole doping, the inflection point at 162 to 174 K evidencing the previously reported 12 K $T_C$ shift, confirmed by anomalous Hall effect (AHE) [25]. The impact of decreasing chemical doping is shown in Figs. 1(d,c), for $x = 0.22$ and 0.15. Progressively insulating behavior is observed, as expected, but *without* a $V_g$-induced IMT. At $x = 0.15$, for example, insulating $\rho(T)$ occurs, but application of -4 V, while generating a ten-fold low $T$ resistivity decrease, is incapable of inducing percolation. As illustrated in Fig. 1(e,f,g), fixing $x = 0.50$ and reducing $t_{exp}$ (from 8 to 5 u.c.) is far more effective in tuning to the brink of percolation. In particular, at 6 u.c., initially insulating $\rho(T)$ occurs, but with application of -4 V driving a ten-fold decrease in low $T$ resistivity, to a state with positive $d\rho/dT$. This is shown more clearly (on a linear $\rho$ scale) in Fig. 2(a), Supplemental Material Sec. C (Fig. S2) [43] confirming finite $T \rightarrow 0$ conductivity at finite $V_g$. A gate-induced percolative IMT is thus realized in LSCO EDLTs, *via* the predicted route of thickness tuning.

Fig. 1(b) shows that these results are at least qualitatively consistent with our theory. The colored points here (right axis) show the *achieved* electrostatic doping ($\Delta s_{exp}$, per Co) as a function of $x_{eff}$, for various experimental thicknesses, $t_{exp}$. As detailed in Supplemental Material Sec. D [43], $x_{eff}$ and $\Delta s_{exp}$ are determined by comparison to LSCO single crystal $\rho(T, x)$ (Figs. S3,S4 [43]). The $x = 0.15$ film in Fig. 1(c), for example, has $x_{eff} = 0.11$ (black point, Fig. 1(b)), the achieved $\Delta s_{exp} =$ 0.11 falling well below the $\Delta s_c = 0.33$ required to reach percolation, consistent with the $V_g$ dependence in Fig. 1(c). The situation is very different for $x = 0.50$, however. The determined $x_{eff}$ here, in the thick-film-limit, is 0.22, consistent with the metallic $\rho(T)$ (Fig. 1(e)). Although quantitative comparisons between theoretical and experimental thickness dependences are complicated by dead layers (Supplemental Material Sec. E, Fig. S5 [43]), decreasing $t_{exp}$ from 8 to 6 u.c. does significantly increase $\Delta s_{exp}$ (to 0.18, Fig. 1(b)), realizing $\Delta s_{exp} > \Delta s_c$, even when comparing to the extreme limit of 2 u.c. (red line).

Importantly, Fig. 2 demonstrates that the gate-induced IMT in these 6 u.c. LSCO films also drives a transition from a short-range magnetically-clustered state to a long-range FM metal. Initial evidence comes from Fig. 2(a), where positive $d\rho/dT$ emerges at $V_g \leq -3$ V, accompanied by inflection near 150 K; in LSCO, such inflection points strongly suggest FM order [25,38]. More direct evidence comes from the AHE, as in Fig. 2(f). Shown here is the 5 K magnetic flux density ($B$) dependence of the transverse conductivity ($\sigma_{xy} = \rho_{xy}/\rho_{xx}^2$, where $\rho_{xy}$ is transverse resistivity and $\rho_{xx}$ is the $B = 0$ longitudinal $\rho$), revealing a remarkable evolution with $V_g$. At $V_g = 0$ and -1 V no AHE is detected, but at $V_g = -2$ V weak AHE emerges, growing into a large, hysteretic effect by -3, -4 V. *This is strong evidence for long-range FM, electrostatically-induced from a non-FM starting point*. Notably, the gate-induced FM is robust, exhibiting 1 T coercivity, remnance of 60% of saturation, and strong perpendicular magnetic anisotropy. The latter is a feature of LSCO under compressive strain, the large anomalous Hall conductivity leading to $\sigma_{xy}(B)$ dominated by magnetism [25]. Setting a small out-of-plane $B = 0.02$ T and measuring $\sigma_{xy}(T)$ then enables an order parameter measurement. As shown in Fig. 2(c), $\sigma_{xy}(T)$ reveals negligible FM at 0 and -1 V, a minor increase at the lowest $T$ at -2 V, but strong FM order at -3, -4 V. The low $T$ downturn in $\sigma_{xy}(T)$ may reflect the upturn in $\rho(T)$ (Fig. 2(a)), or $T$-dependent competition between in-plane and perpendicular anisotropy [25]. Regardless, order-parameter behavior occurs at high $T$, directly demonstrating $T_C$ up to 150 K at -4 V. The electrostatically-induced $T_C$ shift by thickness-tuning to the brink of percolation in LSCO is thus 150 K, over an order of magnitude above the previous 12 K [25].

Magnetoresistance (MR) measurements support these conclusions. Shown in Fig. 2(b) is the $T$ dependence of MR = $[\rho(T,B)-\rho(T,B_c)]/\rho(T,B_c)$, where $B_c$ is the coercive field, and $B$ is fixed, out-of-plane, at 9 T. At $V_g = 0$, the MR magnitude simply increases monotonically on cooling, reaching -30 % at low $T$. As shown in Fig. 2(d,e), this is due to an isotropic, negative, hysteretic MR, with peaks at $\pm B_c$. This is well-known in LSCO, arising due to spin-dependent inter-cluster transport on the insulating side of the IMT, *i.e.*, intergranular giant magnetoresistance [37,38]. This is therefore exactly as expected in a sub-percolative starting film. As the magnitude of $V_g$ is increased, however, this low $T$ isotropic MR is weakened, while a high $T$ MR turns on, around the induced $T_C$ (Fig. 2(b)). This is also typical for LSCO, arising due to the spin-disorder MR known to exist around $T_C$ in the FM metallic phase [37,38]. As shown in Fig. 2(d,e), the low $T$ MR in the gate-induced FM metallic state additionally becomes more anisotropic than at low $V_g$ (by a factor of ~3), due to the onset of anisotropic magnetoresistance (AMR) [46]. The $V_g$ dependence of MR($B,T$) is therefore in excellent agreement with a gate-induced transition from a magnetically-clustered insulator to a long-range FM metal.

A succinct summary of the evolution with $V_g$ is provided in Fig. 2(g,h,i). As $V_g$ decreases from 0 to -4 V the low $T$ resistivity falls by over an order of magnitude, driven by electrostatic hole accumulation (Fig. 2(g)). $T$-dependent measurements confirm this is due to a percolation IMT. Accompanying the IMT, the MR evolves from a state dominated by isotropic low $T$ inter-cluster MR (Fig. 2(h), left axis), to a state with substantial MR near $T_C$ due to field-induced suppression of spin-disorder (Fig. 2(h), right axis), also exhibiting low $T$ AMR (Fig. 2(d,e)). Finally, and most directly, as $V_g$ is decreased below -2 V strong AHE turns on (Fig. 2(i), left axis), the deduced $T_C$ increasing from 0 to 150 K.

While transport evidence for gate-induced percolation to an FM state is strong, *in operando* PNR was also performed, seeking confirmation of long-range FM, as well as the depth-profile of the induced magnetization, $M$. The latter is important, as our recent theory predicts anomalously deep penetration of $M$, due to surface-gating-mediated connection of finite clusters in the film interior [40]. Fig. 3(a) shows the specular neutron reflectivity, $R$, *vs*. out-of-plane scattering wavevector ($Q$), for a 6 u.c. $x = 0.5$ LSCO film at $V_g = -3$ V, $T = 30$ K, and $B = 1$ T (in-plane). Shown are the non-spin-flip reflectivities ($R^{++}$ and $R^{--}$), where the "+" and "-" indicate relative polarizations of the incoming and outgoing neutrons. While weak, splitting indeed occurs between $R^{++}$ and $R^{--}$, consistent with long-range FM at -3 V. This is emphasized in Fig. 3(b), which shows the $Q$ dependence of the spin asymmetry, $SA = [(R^{++} - R^{--})/(R^{++} + R^{--})]$, at $V_g = 0$ and -3 V. The $SA$ is negligible at $V_g = 0$, but becomes finite at -3 V, growing monotonically to $SA = 0.1$ at $Q = 0.09$ Å$^{-1}$. Note that the absence of oscillations in $R(Q)$ and $SA(Q)$ in this $Q$ range is expected, due to the low (6 u.c.) thickness.

While the above confirms gate-induced long-range FM, quantitative refinement provides additional insight. The solid line fits in Fig. 3(a,b) are based on simple depth ($z$) profiles for the nuclear and magnetic scattering length density (SLD), as shown in Fig. 3(c). As described in Supplemental Material Sec. F [43], the nuclear SLD is based on expected values for the substrate, LSCO, and ion-gel, with LSCO thickness and roughness of 25 Å (6.4 u.c.) and 7 Å (1.8 u.c.), respectively. The refined *magnetic* SLD at $V_g = 0$ is indeed zero at all $z$, confirming no long-range FM. At $V_g = -3$ V, however, good fits can only be achieved with finite magnetic SLD, the best-fit $M(z)$ being shown in Fig. 3(c) (right axis). Remarkably, $M$ is quite uniform with depth, the maximum value being 0.34 $\mu_B$/Co, and the magnetic and nuclear roughnesses being identical. To put this in context, at $T_C \approx 150$ K, bulk LSCO has $M \approx 0.8$ $\mu_B$/Co [36]. This LSCO film has perpendicular anisotropy, however, which is not entirely overcome by the available in-plane $B = 1$ T; we thus expect $M < 0.8$ $\mu_B$/Co, consistent with experiment. Importantly, and as elaborated in Supplemental Material Sec. G (Fig. S6) [43], alternative $M(z)$ profiles weighted towards the LSCO surface can be excluded. Our best-fit $M(z)$ in fact extends significantly deeper than the induced carrier profile from Thomas-Fermi calculations, which indicate 90 % carrier confinement in the top 2.5 u.c. This result further validates our recent percolation theory, occurring due to gate-mediated connection of existing finite clusters that penetrate the film thickness, *i.e.*, surface-assisted bulk percolation (Fig. 1(a)) [40]. Importantly, this demonstrates that electrostatic modulation of magnetism need *not* be confined to extreme surface regions.

In summary, using ion-gel-based epitaxial LSCO EDLTs, thorough verification of a recent prediction of the efficacy of thickness tuning to approach the verge of a percolation IMT has been achieved. At an optimal thickness of 6 u.c., a gate-induced transition from an insulating magnetically-clustered state to a long-range FM metal is demonstrated by transport and PNR. This enables giant *electrostatic* $T_C$ modulation of 150 K, dramatically increased over the prior 12 K. This 150 K shift is, to our knowledge, the largest *electrostatic* value in any electrolyte-gated material, and the largest unambiguous $T_C$ shift in a complex oxide by any form of electrostatic gating [47-50]. Our work thus brings electrostatically-induced $T_C$ shifts into the same realm as electrochemical $T_C$ shifts, but with potential advantages in speed and reversibility. Future efforts with higher $T_C$ materials could even realize such $T_C$ shifts around room temperature, creating FMs with electrically-tunable thermal stability.

**Acknowledgments:** Work primarily supported by the National Science Foundation through the UMN MRSEC under DMR-1420013. Partial support (specifically for neutron scattering) is acknowledged from the DOE through the UMN Center for Quantum Materials, under DE-FG02-06ER46275 and DE-SC-0016371. A portion of this research used resources at the Spallation Neutron Source, a DOE Office of Science User Facility operated by Oak Ridge National Lab. Parts of this work were performed in the Characterization Facility, UMN, which receives partial support from NSF through the MRSEC program.

**FIGURE CAPTIONS**

**Fig. 1.** (a) $La_{1-x}Sr_xCoO_{3-\delta}$ (LSCO) EDLT schematic. S/D represents source/drain, $V_g/V_{SD}$ the gate/source-drain voltages, red/blue charges the ion gel cations/anions, and yellow/gray charges electrons/holes, respectively. The LSCO film has thickness $t$ and finite clusters are shown in green. (b) Solid curves (color coded for different theoretical thickness, $t$) show the theoretical surface charge density required to induce percolation, $\Delta s_c$ (left axis), vs. bulk chemical doping, $x_{eff}$. These are obtained from Ref. [40], by rescaling to the LSCO experimental percolation threshold, $x_{c,LSCO}$ [42]. Data points (right axis, color coded to the experimental thickness, $t_{exp}$) show the maximum experimental surface charge density achieved, $\Delta s_{exp}$. Determination of $\Delta s_{exp}$ and $x_{eff}$ is discussed in Supplemental Material Sec. D [43]. The shaded region is discussed in the text. (c-g) Temperature, $T$, dependence of resistivity, $\rho$ (log scales), for LSCO films with nominal $x = 0.15$, 0.22, 0.5, 0.5, and 0.5 at $t_{exp} = 12$, 12, 8, 6, and 5 unit cells, respectively, at $V_g = 0$ to -4 V.

**Fig. 2.** (a,b,c) Temperature, $T$, dependence of (a) zero magnetic field resistivity, $\rho$, (b) 9 T out-of-plane magnetoresistance, MR, and (c) low field (out-of-plane field, $B_{OP} = 0.02$ T) transverse conductivity, $\sigma_{xy}$, at gate bias, $V_g = 0$ to -4 V. (d,e,f) 5 K $B$ dependence of (d) out-of-plane MR, (e) in-plane MR (with current, $I \parallel B$), and (f) $\sigma_{xy}$, at $V_g = 0$ to -4 V. (g,h,i) $V_g$ dependence of (g) $\rho$ at 5 K, (h) out-of-plane MR at 5 K (left axis) and 170 K (right axis), and (i) $\sigma_{xy}$ at 5 K and $B_{OP} = 9$ T (left axis ), and the Curie temperature, $T_C$ (right axis). All data are from the $x = 0.5$, $t_{exp} = 6$ unit cell $La_{1-x}Sr_xCoO_{3-\delta}$ film in Fig. 1(f).

**Fig. 3.** (a) Neutron reflectivity, $R$, vs. scattering wavevector magnitude, $Q$, from an $x = 0.5$, $t_{exp} = 6$ unit cell $La_{1-x}Sr_xCoO_{3-\delta}$ film at gate bias, $V_g = -3$ V, 30 K, and 1 T (in-plane). Black and red

denote non-spin-flip "$R^{++}$" and "$R^{--}$" channels, for both data (points) and fits (solid lines). (b) Spin asymmetry *vs.* $Q$ for $V_g$ = 0, -3 V. Lines are fits with the depth profiles in (c). (c) Depth profiles of the nuclear scattering length density, SLD, (left axis) and magnetization, $M$, (right axis) for $V_g$ = 0, -3 V; the film/substrate interface is at $z$ = 0.

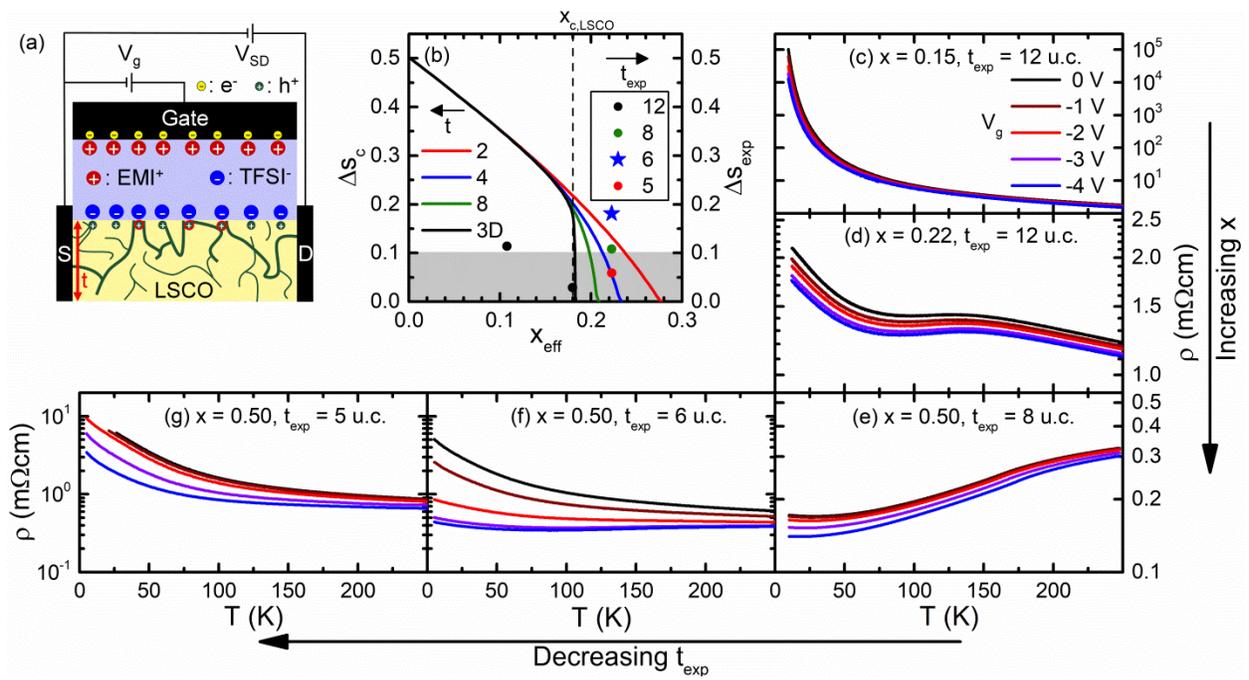

**Figure 1**

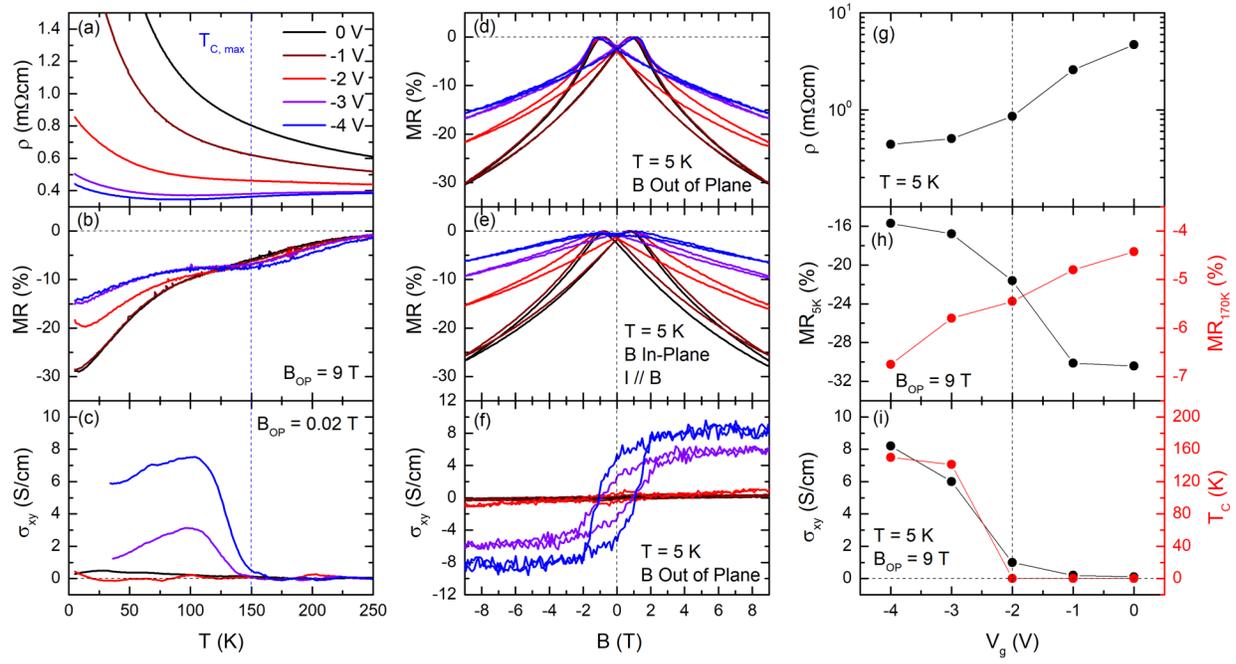

**Figure 2**

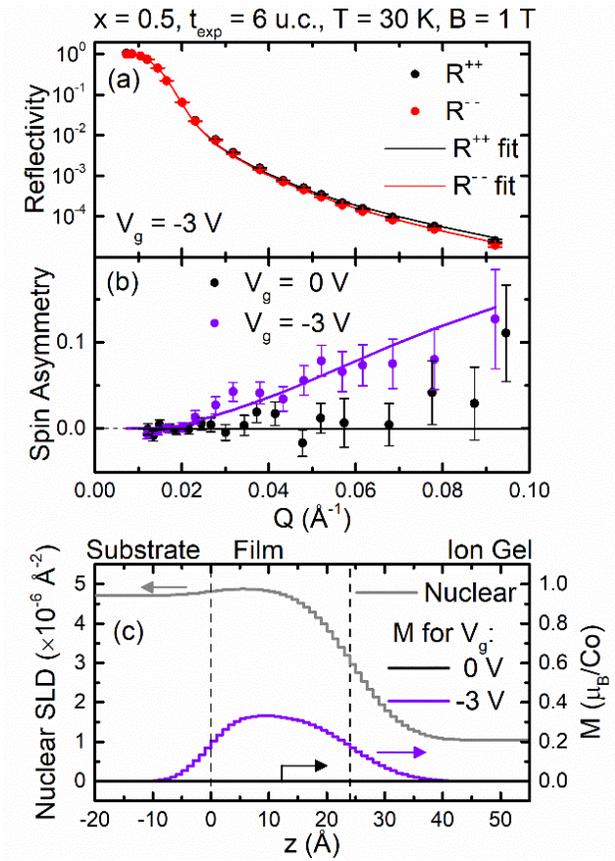

**Figure 3**